\documentclass[aps,prd,superscriptaddress,twocolumn,groupedaddress,amssymb,nofootinbib,10pt]{revtex4}
\pdfoutput=1
\usepackage[pdftex]{graphicx}
\DeclareGraphicsExtensions{.pdf,.jpeg,.png}
\usepackage{bm}
\usepackage{color}
\usepackage{dcolumn}
\usepackage[spanish,english]{babel}
\usepackage{latexsym,float}
\usepackage{epsfig,graphics}
\usepackage{epstopdf}
\usepackage{amssymb}
\usepackage{amsmath}
\usepackage{verbatim}
\usepackage{multirow}
\usepackage{color}
\usepackage{tikz}
\usepackage[utf8]{inputenc}
\usetikzlibrary{matrix}
\usepackage{float}
\DeclareMathAlphabet{\mathpzc}{OT1}{pzc}{m}{it}

\newcommand{\nn}{\nonumber}
\newcommand{\na}{\nabla}
\newcommand{\D}{\partial}

\begin{document}
\title{On gravitational wave modes in Nash theory of gravity }

\author{Phongpichit Channuie}
\affiliation{College of Graduate Studies, Walailak University, Thasala, Nakhon Si Thammarat, 80160, Thailand}
\affiliation{School of Science, Walailak University, Thasala, \\Nakhon Si Thammarat, 80160, Thailand}

\author{Davood Momeni }
\affiliation{Center for Space Research, North-West University, Mafikeng, South Africa }
\affiliation{Tomsk State Pedagogical University, TSPU, 634061 Tomsk, Russia
}
\affiliation{Department of Physics, College of Science, Sultan Qaboos University,
	\\P.O. Box 36, P.C. 123, Al-Khodh, Muscat, Sultanate of Oman}

\author{Mudhahir Al Ajmi }
\affiliation{ Department of Physics, College of Science, Sultan Qaboos University, P.O. Box 36, Al-Khodh 123, Muscat, Sultanate of Oman }

\date{\today}
\begin{abstract}
In this Letter, we consider original Nash theory of gravity. In terms of the scalar fields representation, Nash gravity is equivalent to the action of bi-scalar tensor gravity with four derivative terms of metric tensor. We then quantify the ghost in the theory. In order to study the gravitational wave modes of the theory, we perform a small perturbation over a fixed Minkowski background. We discover the standard wave equation and obtain the solutions as those of the standard general relativity. In order to satisfy  the GR in the weak field limit, we further modify the original Nash theory by adding the Einstein-Hibert term called modified Nash gravity. In this theory, we once study the gravitational wave modes. We derive equations of motion and examine the solutions. We discover the two-mode solutions: the massless graviton and the massive one. Using a uniform prior probability on the graviton mass, we can constrain mass parameter. 

\end{abstract}


\maketitle

\section*{Introduction}
\label{into}
The recent detection of gravitational waves
(GWs) from mergers of primordial black holes (BHs) or
neutron stars (NSs) announced by LIGO/VIRGO \cite{Abbott:2016nmj,Abbott:2017vtc,Abbott:2017oio,Abbott:2016blz,TheLIGOScientific:2017qsa} has pinned
the new era of gravitational wave cosmology. This is a new window to probe the strong gravity physics. However, these individual two-body sources are just one of the many signatures in nature we have already detected. The GWs could be produced by other physical circumstances, for instance, during inflation \cite{Rubakov:1982df} and reheating \cite{GarciaBellido:2007af} or another exotic post-inflationary physics such as phase transitions \cite{Weir:2017wfa,Grojean:2006bp} or topological defects \cite{Vilenkin:2000jqa}. We hope that these distinct backgrounds could be detected by present and future experiments.

As mentioned in Ref.\cite{GarciaBellido:2007af}, a significant fraction of energy in the form of a stochastic background of gravitational waves can be produced during reheating stage. During the process, the amplitude of gravity waves exponentially grow due to the tachyonic preheating. Then a burst of gravitational radiation occurs followed by the end of gravitational waves production. In addition, cosmic defects are products of a phase transition in the early universe. The sources of GW can be of different natures, not only from quantum vacuum fluctuations in the early universe but also  from astronomical sources, e.g. black hole binaries, supernovae, and pulsars. These different sources have their own GW characteristics. In the present work, we consider original Nash theory of gravity and examine possible gravitational wave modes of the theory.

\section*{Nash  gravity and Scalar fields reduction}
We start our study by considering the action of Nash gravity in four-dimensional spacetime, where the metric is a Riemanninan metric $g_{\mu\nu}$. Here the action consists of 
higher order corrected gravity and we adapted a notation as  $g_{\mu\nu}=(-1,1,1,1),~~ R_{\mu\nu}=R^{\alpha}_{\mu\alpha\nu}$ and all Greek indices are spacetime indices running from 0 to 3. The original Nash action takes the form \cite{Nash}
\begin{eqnarray}\label{SN1}
S_{\rm Nash}=\frac{1}{2}\int
d^4x\sqrt{-g} \Big(2R_{\mu\nu}R^{\mu\nu}-R^2\Big).
\end{eqnarray}
In the above action, we have used natural units with $c=1$ and set the gravitational coupling constant $\kappa^2=8\pi G/c^{4}\equiv 1$. Notice that cosmology of Nash gravity was recently investigated by Refs.\cite{Channuie:2018kfm,Channuie:2018now,adne,Lake}. Here the authors of Ref.\cite{Channuie:2018kfm} employed the Noether symmetry technique to quantify exact solutions. Recently, the matter field contents were added to the original theory \cite{Channuie:2018now}.

We rewrite the action of theory, Eq.(\ref{SN1}), using a pair of auxuliary fields $\phi_1,\phi_2$ in the following form
\begin{eqnarray}\label{S2}
S=\frac{1}{2}\int
d^4x&\sqrt{-g}&\Big(2R_{\mu\nu}R^{\mu\nu}-R^2+F_1(R-\phi_1)\nonumber\\&&+F_2(R_{\mu\nu}R^{\mu\nu}-\phi_2)
\Big).
\end{eqnarray}
Here Lagrange multipliers $F_i,i=1,2$ can be quantified by varying the total action with respect to $R$ and $R_{\mu\nu}R^{\mu\nu}$ as variational variables. Employing this technique, we obtain
\begin{eqnarray}\label{S}
F_1=\frac{\partial S}{\partial R}=2R,\ \ F_2=\frac{\partial S}{\partial (R_{\mu\nu}R^{\mu\nu})}=-2.
\end{eqnarray}
Interestingly, substituting the above expressions in Eq. (\ref{S2}) we obtain:
\begin{eqnarray}\label{S3}
S=\frac{1}{2}\int
d^4x\sqrt{-g} \Big(R^2-2R\phi_1+2\phi_2)
\Big).
\end{eqnarray}
What we observe regarding this reduced form of the Nash action is that the matrix $\frac{\partial^2 F_i}{\partial \phi_1\partial \phi_2}=0$ is degenerate . As a result, Nash gravity is equivalent to the action of bi-scalar-tensor gravity with four derivative terms of metric tensor. 
It is illustrative to present another reduced form of Nash gravity using Gauss-Bonnet invariance:
\begin{eqnarray}
&\frac{\delta}{\delta g^{\mu\nu}}&\int\sqrt{-g}d^4x\Big[R_{\alpha\beta\mu\nu}R^{\alpha\beta\mu\nu}-4R_{\mu\nu}R^{\mu\nu}+R^2
\Big]\\&&\nonumber=\frac{\delta}{\delta g^{\mu\nu}}\int\sqrt{-g}d^4xC_{\alpha\beta\mu\nu}C^{\alpha\beta\mu\nu}-\frac{\delta S_{Nash}}{\delta g^{\mu\nu}},
\end{eqnarray}
where $C_{\alpha\beta\mu\nu}$ is the Weyl tensor. A possible alternative form for the action is given as follows:
\begin{eqnarray}
S&=&\int\sqrt{-g}d^4x\Big[2\phi_2-\phi_1^2+F_1 (R-\phi_1)\nonumber\\&&+\frac{1}{2}F_2C_{\alpha\beta\mu\nu}C^{\alpha\beta\mu\nu}+F_2(R_{\mu\nu}R^{\mu\nu}-\phi_2)
\Big)
\Big],
\end{eqnarray}
where again we can fix $F_i$ by varying the total action with respect to $R$ and $R_{\mu\nu}R^{\mu\nu}$. Then the action can be reduced as:
\begin{eqnarray}
S &=&\int\sqrt{-g}d^4x \Big[4\phi_2-\phi_1(\phi_1+2R)+2R^2\nonumber\\&&\quad\quad\quad\quad\quad-2R_{\mu\nu}R^{\mu\nu}
\Big].
\end{eqnarray}

\section*{Looking for ghosts}
In order to investigate untamed ghosts, we limit our investigations to excitations around a vacuum state where there is a potential for an exact solution to the field equations with constant Ricci scalar $R=R_0$. It is adequate to define a reduced Ricci tensor $S_{\mu\nu}$ via the following transformation:
\begin{eqnarray}
S_{\mu\nu}=R_{\mu\nu}-\frac{1}{4}Rg_{\mu\nu}. 
\end{eqnarray}
If we denote the Nash Lagrangian by $L=2R_{\mu\nu}R^{\mu\nu}-R^2$, we can expand the Nash gravity action Eq. (\ref{SN1}) around this vacuum in Taylor series form:
\begin{eqnarray}\label{sghost}
&&
S=\int \sqrt{-g}d^4x\Big[L_0+R^2-R_0^2+\frac{1}{2}\frac{\partial^2 L}{\partial S^2}|_{0}S^2
\Big].
\end{eqnarray}
In the above equation, $S$ is trace of the reduced Ricci tensor and subscript $0$ means to evaluate the expression at this special vacuum solution. It is easy to show that the total action will reduce to an Einstein-Hilbert action with a cosmological constant (first and second terms in action) plus a Ricci squared term, a higher order Weyl squared term and a Gauss-Bonnet term. We can write the final form for (\ref{sghost}) in the following compact form 
\begin{eqnarray}\label{sghost2}
S=\int \sqrt{-g}d^4x\Big[R-2\Lambda+\frac{R^2}{6m_0^2}-\frac{C^2}{2m_2^2}
\Big].
\end{eqnarray}
where $m_0^2$ is a massive spin 0 field while $m_2^2$ corresponds to a massive spin 2, and has a wrong sign of kinetic term and thus has negative energy. As a result we need to treat it as a possible Weyl ghost field. 

\section*{Unstable vacuum in higher order corrected action}
Introducing an auxiliary field $\phi$ in action Eq. (\ref{sghost2}), we obtain:
\begin{eqnarray}\label{sghost3}
&&
S=\int \sqrt{-g}d^4x\Big[(1+\frac{\phi}{3m_0^2})R-\frac{\phi^2}{6m_0^2}-\frac{C^2}{2m_2^2}
\Big].
\end{eqnarray}
Changing the conformal transformation from $g_{\mu\nu}$ to $\tilde{g}_{\mu\nu}=(1+\frac{\phi}{3m_0^2})g_{\mu\nu}$ and finally using another scalar field $\tilde{\phi}=\ln(1+\frac{\phi}{3m_0^2})$ we find another alternative form for action (\ref{sghost3}) in Einstein frame:
\begin{eqnarray}\label{sghost4}
S&=&\int \sqrt{-g}d^4x\Big[\tilde{R}-\frac{3}{2}(\tilde{\nabla}\varphi)^2-\frac{3}{2}m_0^2(1-e^{-\varphi})^2\nonumber\\&&\quad\quad\quad\quad\quad-\frac{\tilde{C}^2}{2m_2^2}
\Big].
\end{eqnarray}
The model possesses an instability in vacuum as demonstrated in a general model of modified gravity theories \cite{Chiba:2005nz}. 

\section*{Graviton propagation}
\label{gw}
In the absence of matter field Lagrangian, we can derive the equation of motion (EoM) of gravitational sector. This can be done by varying the action given in Eq.(\ref{SN1}) with respect to the metric $g_{\mu\nu}$  in the metric formalism to obtain
\begin{eqnarray}
-2RG_{\mu\nu}&=&\frac{1}{2}g_{\mu\nu}\left(2R_{\mu\nu}R^{\mu\nu}+R^2\right)-2(g_{\mu\nu}\Box-\na_\mu\na_\nu)R\nn\\&&
-4R^a_\mu
R_{a\nu} -2
g_{\mu\nu}\na_\alpha\na_\beta R^{\alpha\beta}\nonumber\\&&-2\Box 
R_{\mu\nu}+4\na_\alpha\na_\beta~R^\alpha_{~(\mu}\delta^\beta_{~\nu)}.\label{fieldeqs}
\end{eqnarray}
Notice that we observe a forth order EoM for metric components and second order for Ricci tensors. Here $\Box=g^{\alpha\beta}\na_\alpha\na_\beta$ is the usual d'Alembert operator. For simplicity, we have used symmetrized tensor representation where  $T_{(\alpha\beta)}=\frac{1}{2}(T_{\alpha\beta}+T_{\beta\alpha})$. This denotes symmetrization with respect to the pair of the indices $(\alpha,\beta)$. In this work we attempt to investigate gravitational wave modes in the Nash theory of gravity. Here we are going to derive the EoM for scalar $R$. Taking the trace of Eq.(\ref{fieldeqs}), we obtain
\begin{eqnarray}
\Box R=\frac{1}{2}g^{\alpha\beta}\na_\alpha\na_\beta R. \label{trace1}
\end{eqnarray}
 Note that Eq.(\ref{trace1}) is a Klein-Gordon equation for the scalar field $\Phi$:
\begin{eqnarray} 
\Box \Phi = \frac{dV}{d\Phi} .
\end{eqnarray}
Here we have defined $\Phi \equiv
R \,\, {\rm and} \,\, dV/d\Phi \equiv {\rm RHS ~of~ (\ref{trace1})}$.
\subsection*{Linearization around Minkowski $\mathcal{M}_4$}
In order to examine GW modes of the model, we need to perturb the above field equation over a fixed Minkowski background $\eta_{\mu\nu}={\rm dia}(-,+,+,+)$. Here we expand the metric and scalar field in the following form:
\begin{eqnarray} 
g_{\mu\nu}&=&\eta_{\mu\nu}+h_{\mu\nu}\,\,, \nn \\
\Phi&=&\Phi_0+\delta \Phi .
\end{eqnarray}
A first order perturbation on the Ricci scalar can be denoted by $\delta \Phi=\delta R$. Furthermore, the perturbation on the Riemann and Ricci tensors take the following form:
\begin{eqnarray} 
\delta R_{\mu\nu} &=& \frac{1}{2}\left(\D_\sigma \D_\nu
h^\sigma_{~\mu}+\D_\sigma \D_\mu h^\sigma_{~\nu}-\D_\mu \D_\nu
h-\Box h_{\mu \nu} \right),\nn\\
\delta R &=& \D_\mu \D_\nu h^{\mu \nu}-\Box h\nn,
\end{eqnarray}
where $h=\eta^{\mu \nu} h_{\mu \nu}$. From Eq.(\ref{trace1}), we obtain the Klein-Gordon equation for the scalar perturbation $\delta \Phi$ as follows:
\begin{eqnarray} 
\Box \delta \Phi=R_0\delta \Phi = m_{spin0}^2 \delta \Phi\,.
\label{kgordon1}
\end{eqnarray}
Note that $R_0=R(\eta_{\mu\nu})=0$. We then perturb the field equations (\ref{fieldeqs}) to obtain
\begin{eqnarray} 
&&\frac{1}{3}(\eta_{\mu\nu}\Box -\D_\mu\D_\nu)\delta
R-2\eta_{\mu\nu} \D_a\D_b \delta{R}^{ab}-2\Box(
\delta{R}_{\mu\nu})\nonumber\\&&+2
\D_a\D_b~\delta{R}^a_{~(\mu}\delta^b_{~\nu)}=0 .
\end{eqnarray} 
Commonly, it is more convenient to work in the Fourier space. In this case we just replace $\D_\gamma
h_{\mu\nu}\rightarrow i k_\gamma h_{\mu\nu}$ and $\Box h_{\mu\nu}
\rightarrow -k^2 h_{\mu\nu}$. Then the above equation becomes
 \begin{eqnarray} 
&&\frac{1}{3}(\eta_{\mu\nu}k^2 -k_\mu k_\nu)\delta
R+2\eta_{\mu\nu} k_a k_b  \delta{R}^{ab}+2k^2
\delta{R}_{\mu\nu}\nonumber\\&&-4 k_a
k_b~\delta{R}^a_{~(\mu}\delta^b_{~\nu)}=0.\label{fields2}
\end{eqnarray} 
In the case of the metric perturbation, we write
\begin{eqnarray}  
h_{\mu\nu}=\bar{h}_{\mu\nu}-\frac{\bar{h}}{2}~
\eta_{\mu\nu}+\eta_{\mu\nu} h_f, \label{gauge}
\end{eqnarray} 
where $h_{f}$ denotes $h-\bar h$. Our gauge freedom takes the usual conditions $\D_\mu\bar{h}^{\mu\nu} =0$ and $\bar{h}=0$. The first of these
conditions implies that $k_\mu \bar{h}^{\mu\nu} =0$ while the
second dictates 
\begin{eqnarray} 
h_{\mu\nu}&=&\bar{h}_{\mu\nu}+\eta_{\mu\nu} h_f, \\
h&=&4 h_f.
\end{eqnarray}
With these perturbations, we find \cite{Bogdanos:2009tn}:
\begin{eqnarray}  
 \delta
R_{\mu\nu}&=&\frac{1}{2}\left(2k_\mu k_\nu h_f+k^2 \eta _{\mu\nu}
h_f+k^2 \bar{h}_{\mu\nu}\right), \nn\\
\delta R &=& 3k^2 h_f,\nn\\
k_\alpha k_\beta ~\delta
R^{\alpha~~~~~\beta}_{~~(\mu\nu)~}&=&-\frac{1}{2}\left((k^4
\eta_{\mu\nu}-k^2 k_\mu k_\nu)h_f+k^4 \bar{h}_{\mu\nu}\right),\nn\\
k_a k_b~\delta{R}^a_{~(\mu}\delta^b_{~\nu)}&=&\frac{3}{2}k^2k_\mu
k_\nu h_f\,. \label{results1}
\end{eqnarray} 
Substituting (\ref{gauge})-(\ref{results1}) into (\ref{fields2}) and performing some algebraic simplification, we obtain 
\begin{eqnarray} 
k^4\bar{h}_{\mu\nu}=
(\eta_{\mu\nu}k^2 -k_\mu k_\nu)\delta R,
\end{eqnarray} 
where we have used $R_0=0$. The above relation implies
\begin{eqnarray} 
k^4\bar{h}_{\mu\nu}=0\,,
\label{solution} 
\end{eqnarray} 
where we have defined $m^2_{\rm spin\,2}=0$. Note that from Eq.(\ref{solution}) we have a modified
dispersion relation which corresponds to a massless spin-2 field
($k^2=0$) and a massless spin-2 ghost mode
$k^2=0$ . The solution to Eq.(\ref{solution}) can be parametrized by plane waves:
\begin{eqnarray} 
\bar{h}_{\mu\nu}&=&A_{\mu\nu}
(\vec{p}) \cdot  \exp(ik^\alpha x_\alpha)+c.c. ,\label{pw1}
\end{eqnarray} 
where
\begin{eqnarray}
k^{\alpha}\equiv(\omega_{m_{spin2}},\vec{p}),\ \ \omega_{m_{spin2}}=p.\label{eq:keq}
\end{eqnarray} 
Here $m_{spin2}$ is zero in the case of massless
spin-2 mode and the polarization tensors is defined as $A_{\mu\nu}
(\vec{p})$. Notice that we obtain the standard wave equation (\ref{solution}) and solutions (\ref{pw1}) in general relativity.

\section*{Modified Nash gravity with Einstein-Hilbert term}
The original Nash theory doesn't recover GR results in the weak limit. Actually it will be very useful as a GR complementary part at UV regime and consequently very good as a potentially candidate for quantum gravity. Regarding the recent observations of GW, we are investigating a soft modification of the Nash action by adding an IR completeness term $R$ to the action as following:
\begin{eqnarray}\label{SN}
S_{EH-Nash}=\frac{1}{2}\int
d^4x\sqrt{-g} \Big( \alpha R+2R_{\mu\nu}R^{\mu\nu}-R^2\Big).
\end{eqnarray}
The action is an Einstein-Hilbert corrected Nash gravity. Note that for sake or dimensionally correctness of the action we introduced a coefficient $\alpha$ with dimension $[\alpha]=[R]=L^{-2}=(\mbox{Gev})^2$. It is very straightforward to derive EoM in the metric formalism, 
\begin{eqnarray}\label{eom-mod}
&&(\alpha-2R)G_{\mu\nu}=\frac{1}{2}g_{\mu\nu}\Big(2R_{\mu\nu}R^{\mu\nu}+R^2\Big)\\&&\nonumber-2(g_{\mu\nu}\Box R-\nabla_{\mu}\nabla_{\nu}R)-4R_{\mu}^{\gamma}R_{\gamma\nu}\\&&\nonumber-2g_{\mu\nu}\nabla_{\gamma}\nabla_{\theta}R^{\gamma\theta}-2\Box R_{\mu\nu}+4\nabla_{\gamma}\nabla_{\theta}R^{\gamma}_{(\mu}R^{\theta}_{\theta)}.
\end{eqnarray}
Trace of the above field equation gives
\begin{eqnarray}
&&\Box \phi=\frac{1}{9}\Big( \alpha R-4R_{\mu\nu}R^{\mu\nu}+R^2\Big),
\end{eqnarray}
where $\phi\equiv \alpha-\frac{2}{3}R$. Now we can linearize it around flat spacetime to obtain:
\begin{eqnarray}
&&\Box \delta\phi=\frac{\alpha}{4}\delta\phi=m_s^2\delta\phi.
\end{eqnarray}
For metric perturbations, we write
\begin{eqnarray}
&&\alpha(\delta R_{\mu\nu}-\frac{1}{2}\eta_{\mu\nu}\delta R)=\\&&\nonumber
(\eta_{\mu\nu}k^2-k_{\mu}k_{\nu})(\delta\phi-\frac{4}{3}\delta R)\\&&\nonumber
2\eta_{\mu\nu}k_{\gamma}k_{\theta}\delta R^{\gamma\theta}+2k^2\delta R_{\mu\nu}-4k_{\gamma}k_{\theta}\delta R^{\gamma}_{(\mu}R^{\theta}_{\theta)}.
\end{eqnarray}
Using the same gauge freedom we used in the previous section:
\begin{eqnarray}
&&\partial_{\mu}\bar h^{\mu\nu}=\partial_{\mu}(h^{\mu\nu}-\frac{\bar h}{2}\eta^{\mu\nu}+\eta^{\mu\nu}h_f)=0\\&&\nonumber  \ \  \bar h=0,\ \  h=4h_f,
\end{eqnarray}
we simplify it and find:
\begin{eqnarray}
\Big(k^2+\frac{k^4}{m_{spin2}^2}\Big)\bar h_{\mu\nu}=0,
\end{eqnarray}
where $m_{spin2}^2=-\frac{\alpha}{2}$ and meanwhile we have $\Box h_f=m_s^2h_f$. We deduce that two modes exist, one is massless graviton with dispersion relation $k^2=0$. the other is a massive (if $\alpha>0$) ghost (non -ghost if $\alpha<0$) with dispersion relation $k^2=-m_{spin2}^2=\frac{\alpha}{2}$. 

In case of negative $\alpha$, we constrain a graviton mass using a uniform
prior probability on the graviton mass \cite{TheLIGOScientific:2016src} $m_{g}\subset[10^{-26},10^{-16}]$ to obtain
\begin{eqnarray}
|\alpha|\subset [10^{-52},10^{-32}].
\end{eqnarray}
Notice that due to the extremely small values of $\alpha$ the last two terms play a leading role in the modified theory.

\section*{Summary}
In this work, we considered original Nash theory of gravity. In terms of the scalar fields representation, Nash gravity is equivalent to the action of bi-scalar tensor gravity with four derivative terms of metric tensor. We then quantified the ghost in the theory and found an instability in vacuum of the theory. We studied the graviton propagation by performing a small perturbation over a fixed Minkowski background. We discovered the standard wave equation and obtain the solutions as those of the standard general relativity. 

In order to satisfy the GR in the weak field limit, we further modified the original Nash theory by adding the Einstein-Hibert term leading to the so-called modified Nash gravity. In this theory, we once studied the gravitational wave modes we derived equations of motion and examine the solutions. We discovered the two-mode solutions: the massless graviton and the massive one. Using a uniform prior probability on the graviton mass, we constrained mass parameter to obtain $|\alpha|\subset [10^{-52},10^{-32}]$.

\section{Acknowledgment} 
D. Momeni and M. Al Ajmi would like to acknowledge the support of Sultan Qaboos University under the Internal Grant (IG/SCI/PHYS/19/02).


\end{document}